\documentclass[a4paper, 11pt,3p]{article}
\usepackage{xcolor}
\usepackage{placeins}
\usepackage{mdframed}
\usepackage{geometry}
\usepackage{authblk}
\usepackage{enumitem}
\usepackage{url}
\usepackage{multicol} 
\usepackage{breqn}
\usepackage{amsmath}
\title{\textbf{Evaluating Joule heating’s influence on heat transfer and entropy generation in MHD channel flow: A parametric study and ill-posed problem solution using PINNs}}
\author[1]{E. Ghaderi}
\author[1]{M. Bijarchi\footnote{corresponding author: bijarchi@sharif.edu}}
\author[1]{S. Kazemzadeh Hannani}
\author[1]{A. Nouri-Boroujerdi}
\affil[1]{Department of Mechanical Engineering, Sharif University of Technology, Tehran, Iran}
\usepackage{graphicx}
\usepackage[%
colorlinks=true,
pdfborder={0 0 0},
linkcolor=blue,
citecolor=blue,
]{hyperref}
\allowdisplaybreaks
\begin{document}
	\maketitle
	\setlist[description]{font=\normalfont}
	\vspace{-1.5cm}
	\section*{Abstract}
	In this study the effects of Joule heating parameter on entropy generation and heat transfer in MagnetoHydroDynamic (MHD) flow inside a channel is investigated by means of Physics-Informed Neural Networks (PINNs) in form of a parametric analysis in addition to exploring the solution to the ill-posed problem.
All of the governing equations are reformulated in terms of first order derivatives and the dimensionless form of the governing equations has been employed to further lessen the number of parameters and achieve better compatibility with loss function terms.
Dimensionless groups such as Reynolds number, Prandtl number, Hartmann number and Joule heating parameter have been designated as the input for the neural network in order to perform the parametric study.
Besides achieving high accuracy for case of parameters confined in the predefined ranges, the generalization ability of the method is depicted by solving the problem for the cases where the dimensionless parameters were outside of the assumed ranges.
Moreover, the ability of handling Neumann boundary conditions is also investigated in the present study despite being neglected prevalently in the literature concerning PINNs. 
The effects of Joule heating parameter on entropy generation are researched using a parametric approach utilizing PINNs which is another novel aspect of the article at hand.
As a concluding remark, an ill-posed problem is also defined such that the Joule heating parameter is included in a learning process and the method has been able to determine Joule heating parameter as a parameter of interest alongside other learnable parameters of the neural network such as weights and biases.
\\
\textbf{Keywords:} Joule Heating Parameter, MHD Flow, PINNs, Entropy Generation, Ill-posed Problem
\begin{figure*}[h]
	\centering
	\begin{mdframed}[everyline=true,linewidth=1pt]
		\textbf{Nomenclature}
		\begin{footnotesize}
		\begin{multicols}{2}
			\begin{description}[noitemsep]
				\item[\textbf{Symbols}]
				\item[$B_0$] external magnetic field (T)
				\item[$C_p$] specific heat capacity ($J \: kg^{-1}\:K^{-1}$)
				\item[$H$] channel Height (m)
				\item[$Ha$]Heartmann dimensionless number
				\item[$J$] Joule heating parameter 
				\item[$L$] channel length (m)
				\item[$N$] number of Points
				\item[$p$] pressure (Pa)
				\item[$Pr$] Prandtl Number
				\item[$q$]heat flux components ($W\:m^{-2}$)
				\item[$r$] residuals
				\item[$Re$] Reynolds Number
				\item[$S_{gen}$] local entropy generation ($W\:K^{-1}\:m^{-3}$)
				\item[$T$] temperature (K)
				\item[$u$] velocity Component 
				\item[] in horizontal Direction ($m\:s^{-1}$)
				\item[$v$] velocity Component 
				\item[] in vertical Direction ($m\:s^{-1}$)
				\item[$x,y$] Cartesian coordinates (m)
				\item[\textbf{Greek Letters}]
				\item[$\beta$] boundary conditions loss function coefficient
				\item[$\lambda$] fluid Thermal Conductivity ($W\:m^{-1}\:K^{-1}$)
				\item[$\nu$] kinematic Viscosity ($m^2\:s^{-1}$)
				\item[$\rho$] fluid Density ($kg\:m^{-3}$)
				\item[$\sigma$] electric conductivity of the Fluid ($S\:m^{-1}$)
				\item[$\tau$] Cauchy stress tensor components (Pa)
				\item[\textbf{Superscripts}]
				\item[$\ast$] dimensionless entity
				\item[\textbf{Subscripts}]
				\item[BC] Boundary Conditions
				\item[Con] Continuity equation
				\item[En] Energy equation
				\item[Inlet] Inlet of the channel
				\item[Max] Maximum
				\item[Mom] Momentum Equation
				\item[Out] Outlet of the channel
				\item[PDE] Partial Differential Equation
				\item[Wall] channel Walls
				\item[Total] Total loss function
			\end{description}
		\end{multicols}
		\end{footnotesize}
	\end{mdframed}
\end{figure*}
\FloatBarrier
\pagebreak
\section{Introduction}
Heat transfer enhancement is a matter of importance and of great use. It's importance prompts researchers and engineers to seek methods to further find solutions to improve heat transfer in their research.
Methods such as using extended surfaces and fins\cite{1}, Nano-fluids\cite{2}, using external sources of power in forms such as ElectroHydroDynamics\cite{3} or MagnetoHydroDynamics\cite{4} and Porous medium\cite{5} in addition to other methods have been explored in the literature.\cite{6}
Among the aforementioned methods, using magnetic fields is perceived to be as one of the most effective methods, which was studied by Dulikravich and Lynn\cite{7}, Davidson\cite{8}, Sheikholeslami and Rokni\cite{9}.
One of the major reasons of increased heat transfer in presence of magnetic fields is the variations in velocity due to the magnetic fields. Since the velocity terms are also included in the energy equation, the variations in velocity cause changes in the temperature profile of the flow.
Moreover, the presence of magnetic fields can directly manifest itself as a Joule heating heat source term in the energy equation which would cause internal heat generation in the medium.\cite{10}
Formerly used methods such as Finite Element Method (FEM), Finite Difference Method (FDM) and Finite Volume Method (FVM) were employed by researchers to better study the heat transfer. 
The disadvantages of the mentioned methods include complex grid generation process, {computation} time increase with the growth of dimensions and reduced precision due to using the discrete form of the governing equations. 
Furthermore, in inverse heat transfer problems and ill-posed problems where the fluid properties or the boundary conditions are not known, these conventional methods fall short.\cite{11} 
One of the novel methods explored recently to counter the issues mentioned beforehand, is using Artificial Intelligence (AI), particularly in form of PINNs.
AI has been a new area of science with growing interest in the decade prior, being vastly used in areas such as Natural Language Processing, Computer Vision, Recommender Systems, Self-driving Cars as well as issues previously thought impossible to resolve.\cite{12}
One of the main subsets of artificial intelligence is machine learning, which employs methods involving neural networks with different structures such as Multilayer Perceptron (MLP), Convolutional Neural Network (CNN) and Recurrent Neural Network (RNN).\cite{13}
In conventional methods of developing neural networks, learning process requires a lot of data, which makes them data-driven methods. Unfortunately data is often scarce in {contexts} involving fluid dynamics and heat transfer, and acquiring proper data for training requires complex simulations or costly experiments. 
Moreover, the governing physics of the problem is ignored in data-driven methods, while most phenomenons in fluid dynamics and heat transfer are described using conservation laws and differential equations, making them interesting to pursue by researchers.\cite{11} 
These issues prompted the emergence of Scientific Machine Learning in recent years.\cite{14}
Initially, Raissi et al \cite{15} introduced PINNs and applied machine learning. In this method, differential equations accompanying boundary conditions and initial conditions are imposed as a physical constraint to guide the learning process of the network. 
These constraints improve the accuracy of the neural network, making learning possible when no measured data for the learning process is available.
Raissi et al \cite{16} later studied the equations of fluid dynamics using PINN, pointing out its possible applications regarding problems where obtaining pressure and velocity is difficult such as plane aerodynamics and blood flow in veins.
Applying PINNs to other problems such as flow around a cylinder\cite{17}, flow inside veins subjected to stenosis\cite{18} and turbulent flows\cite{19} alongside applications in heat transfer\cite{20} were hence explored in other articles.
numerous Researches have been conducted to improve the performance of PINNs and reduction of computation time. \cite{21},\cite{22},\cite{23} The recent findings on applying PINNs to various problems yielded promising results, putting the use of this method for analyzing various problems in fluid dynamics and heat transfer in prospect.
An examination of the available literature indicates that because of the novelty of PINNs and the complex nature of the flow under the influence of magnetic field, a conclusive study concerning flow analysis and heat transfer in presence of magnetic fields has not been conducted according to the author's knowledge. 
Furthermore, parametric study and obtaining the physical properties in form of an inverse problem using PINNs for fluid flow in presence of magnetic fields are not explored yet.

In the present study, parametric study of fluid flow and heat transfer in presence of magnetic fields is conducted using PINNs. The effects of dimensionless entities such as Reynolds number, Prandtl number and Hartmann number in addition to Joule heating parameter on pressure drop and heat transfer have been studied with a wide range assigned to each parameter.
The ability of the proposed method to obtain solution in cases where parameters were outside of the predefined intervals is also another novel aspect of this research.
Precision and speed of calculations is enhanced by implementing the dimensionless form of the governing equations in addition to using the low-order form of the governing equations. 
In section concerning the physical properties of the problem, the Joule heating parameter is studied as a fundamental dimensionless parameter in problems involving presence of magnetic fields. 
In this article, Dirichlet boundary conditions as well as Neumann boundary conditions are imposed in addition to the governing equations as physical constraints, and the performance of the proposed method employing PINNs is assessed.
\FloatBarrier
\section{{Problem setup}}
\begin{figure}[h]
	\centering
	\includegraphics{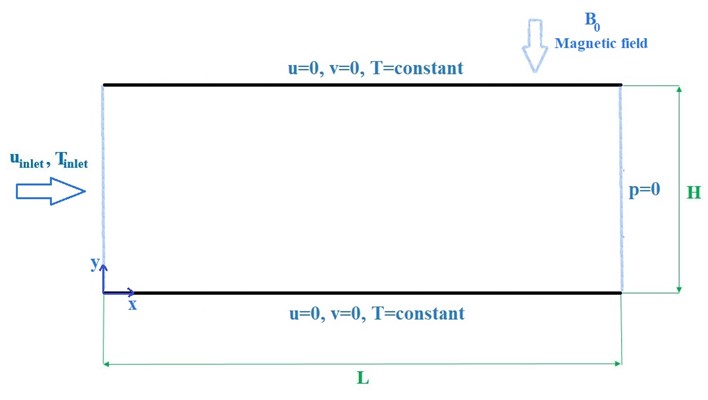}
	\caption{Schematic of the geometry and boundary conditions}
	\label{Fig1}
\end{figure}

In this study, flow inside a channel under the influence of magnetic fields is investigated. Geometry of the domain and boundary conditions are shown in Fig.\ref{Fig1}. In this study, velocity components in the $x$ and $y$ directions are denoted by u and v, respectively. Temperature, pressure, and channel dimensions are also represented by T, p and H, respectively.
 The fluid enters the channel with a given velocity and temperature and its temperature gradually increases as it flows through the channel.
\\
Assuming the principles of continuum mechanics are applicable in this context, differential equations governing the fluid motion under the influence of magnetic field are presented in form of Equations \ref{Eq1} to \ref{Eq4}.\cite{10}
In order to couple pressure and velocity in the mentioned equations, the Poisson equation is utilized instead of continuity equation as depicted in Eq.\ref{Eq3}.\cite{40} 
In order to obtain the governing equations, it is assumed that the fluid is homogeneous, single phase and exhibits Newtonian behavior while having constant properties. The flow is also modeled as incompressible, laminar and steady. Further more, effects of gravity, induced magnetic fields and external electric fields are disregarded in this study.\cite{24} 
Effects of heat transfer by means of radiation and free convection are neglected and magnetic fields are incorporated in form of a source term in the energy equation.\cite{25}
\begin{align}
	&u\frac{\partial u}{\partial x}+v\frac{\partial u}{\partial y}=-\frac{1}{\rho}\frac{\partial p}{\partial x}+\upsilon\left(\frac{\partial^2u}{\partial x^2}+\frac{\partial^2u}{\partial y^2}\right)-\frac{\sigma B_0^2}{\rho}u
	\label{Eq1}
	\\
	&u\frac{\partial v}{\partial x}+v\frac{\partial v}{\partial y}=-\frac{1}{\rho}\frac{\partial p}{\partial y}+\upsilon\left(\frac{\partial^2v}{\partial x^2}+\frac{\partial^2v}{\partial y^2}\right)
	\label{Eq2}
	\\
	&\frac{\partial^2p}{\partial x^2}+\frac{\partial^2p}{\partial y^2}=-\rho\left(\frac{\partial u}{\partial x}\frac{\partial u}{\partial x}+2\frac{\partial u}{\partial y}\frac{\partial v}{\partial x}+\frac{\partial v}{\partial y}\frac{\partial v}{\partial y}\right)
	\label{Eq3}
	\\
	&u\frac{\partial T}{\partial x}+v\frac{\partial T}{\partial y}=\frac{\lambda}{\rho c_p}\left(\frac{\partial^2T}{\partial x^2}+\frac{\partial^2T}{\partial y^2}\right)+\frac{\sigma B_0^2}{\rho c_p}u^2
	\label{Eq4}
\end{align}
In the above equations, $\upsilon$, $\rho$ and $c_p$ symbolize density, kinematic viscosity and specific heat capacity of the fluid {,respectively}. Also, $B_0$, $\sigma$ and $\lambda$ denote external magnetic fields, electric conductivity and thermal conductivity of the fluid respectively. The problem is solved considering two different boundary conditions, namely temperature (Dirichlet boundary condition) and heat flux (Neumann boundary condition), making it possible to assess the performance of the proposed solution employing PINN under these circumstances.
\FloatBarrier
\subsection{\textbf{Solving Procedure}}
In this section, the governing equations and boundary conditions are investigated using PINN method in accordance with the physics introduced in the previous section.
\FloatBarrier
\subsubsection{\textbf{Governing Equations}}
In the studies conducted by Rao et al.\cite{17}, Laubscher\cite{26} and Hu et al.\cite{27} using the low-order form of the derivatives is encouraged when using the PINN method. Consequently, in this study the governing equations are rewritten as per Eq.\ref{Eq5} to \ref{Eq7} in order to reduce the computational cost in the process of computing backward derivatives and improve the implementation of the Neumann boundary conditions (heat flux as investigated in the present article). The governing equations are thus Implemented in form of first-order derivatives as depicted in Equations \ref{Eq5} to \ref{Eq7}.  
Additionally, the presented method is also applicable to problems where the fluid properties are not constant such as where non-Newtonian or non-Fourier models are used.
\begin{align}
	&u\frac{\partial\ u}{\partial\ x}+v\frac{\partial\ u}{\partial\ y}=\frac{1}{\rho}(\frac{\partial\tau_{11}}{\partial\ x}+\frac{\partial\tau_{12}}{\partial\ y})-\frac{\sigma\ B_0^2}{\rho}u
	\label{Eq5}\\
	&u\frac{\partial\ v}{\partial\ x}+v\frac{\partial\ v}{\partial\ y}=\frac{1}{\rho}(\frac{\partial\tau_{12}}{\partial\ x}+\frac{\partial\tau_{22}}{\partial\ y})
	\label{Eq6}\\
	&u\frac{\partial\ T}{\partial\ x}+v\frac{\partial\ T}{\partial\ y}=\frac{1}{\rho\ c_p}(\frac{\partial\ q_x}{\partial\ x}+\frac{\partial\ q_y}{\partial\ y})+\frac{\sigma\ B_0^2}{\rho\ c_p}u^2
	\label{Eq7}
\end{align}
The term $\tau$ in the above equation stands for various 2-D Cauchy stress tensor components which are defined as follows:
\begin{align}
	&\tau_{11}=-p+2\rho\upsilon\frac{\partial\ u}{\partial\ x}\\
	&\tau_{22}=-p+2\rho\upsilon\frac{\partial\ v}{\partial\ y}\\
	&\tau_{12}=\rho\upsilon\left(\frac{\partial\ u}{\partial\ y}+\frac{\partial\ v}{\partial\ x}\right)
\end{align}
Using the above equations, the continuity equation can be simplified using Eq.\ref{Eq11}, relating pressure and Cauchy stress tensor components.
\begin{equation}
	p=-\frac{\tau_{11}+\tau_{22}}{2}
	\label{Eq11}
\end{equation}
According to the Fourier law of heat conduction, $x$ and $y$ components of heat flux are defined as such:\cite{28}
\begin{align}
	&q_x=\lambda\frac{\partial\ T}{\partial\ x}\\
	&q_y=\lambda\frac{\partial\ T}{\partial\ y}
\end{align}
To write these equations in dimensionless form, several dimensionless entities are defined by the following equations:

\begin{align}
	&x^\ast=\frac{x}{H}\\ \nonumber
	&y^\ast=\frac{y}{H}\\ \nonumber
	&u^\ast=\frac{u}{u_{max}}\\ \nonumber
	&v^\ast=\frac{v}{u_{max}}\\ \nonumber
	&p^\ast=\frac{p}{{\rho u_{max}}^2}\\ \nonumber
	&T^\ast=\frac{T}{T_{wall}}\\ \nonumber
	&{\tau_{11}}^\ast\ =\frac{\tau_{11}}{{\rho\ u_{max}}^2}\\ \nonumber
	&{\tau_{22}}^\ast\ =\frac{\tau_{22}}{{\rho\ u_{max}}^2}\\ \nonumber &{\tau_{12}}^\ast\ =\frac{\tau_{12}}{{\rho u_{max}}^2}\\ \nonumber
	&{q_x}^\ast\ =\frac{q_x}{(\lambda\ T_{wall})/H}\\ \nonumber
	&{q_y}^\ast=\frac{q_y}{(\lambda\ T_{wall})/H} \nonumber
\end{align}
In these equations, $U_{max}$ and $T_{wall}$ are defined as the maximum velocity in the center of the parabolic velocity profile (or the inlet velocity of the uniform flow) and the wall temperature, respectively. In this study, it is assumed that the wall temperature exceeds the fluid temperature, thereby heat flows from the wall to the fluid.  
Substituting the mentioned dimensionless entities into the governing equations of the problem results in the following set of equations:
\begin{align}
	&u^\ast\frac{\partial\ u^\ast}{\partial\ x^\ast}+v^\ast\frac{\partial\ u^\ast}{\partial\ y^\ast}=(\frac{\partial{\tau_{11}}^\ast}{\partial\ x^\ast}+\frac{\partial{\tau_{12}}^\ast}{\partial\ y^\ast})-\frac{{Ha}^2}{Re}u^\ast\\
	&u^\ast\frac{\partial\ v^\ast}{\partial\ x^\ast}+v^\ast\frac{\partial\ v^\ast}{\partial\ y^\ast}=(\frac{\partial{\tau_{12}}^\ast}{\partial\ x^\ast}+\frac{\partial{\tau_{22}}^\ast}{\partial\ y^\ast})\\
	&u^\ast\frac{\partial\ T^\ast}{\partial\ x^\ast}+v^\ast\frac{\partial\ T^\ast}{\partial\ y^\ast}=\frac{1}{RePr}\left(\frac{\partial{q_x}^\ast}{\partial\ x^\ast}+\frac{\partial{q_y}^\ast}{\partial\ y^\ast}\right)+J{u^\ast}^2\\
	&{\tau_{11}}^\ast\ =-p^\ast+\frac{2}{Re}\frac{\partial\ u^\ast}{\partial\ x^\ast}\\
	&{\tau_{22}}^\ast\ =-p^\ast+\frac{2}{Re}\frac{\partial\ v^\ast}{\partial\ y^\ast}\\
	&{\tau_{12}}^\ast\ =\frac{1}{Re}(\frac{\partial\ u^\ast}{\partial\ y^\ast}+\frac{\partial\ v^\ast}{\partial\ x^\ast})\\
	&p^\ast=-\frac{{\tau_{11}}^\ast+{\tau_{22}}^\ast}{2}\\
	&{q_x}^\ast=\frac{\partial\ T^\ast}{\partial\ x^\ast}\\
	&{q_y}^\ast=\frac{\partial\ T^\ast}{\partial\ y^\ast}
\end{align}
As the equations are written in dimensionless form, several dimensionless parameters such as Reynolds number (Re=$\frac{u_{max}H}{\upsilon}$), Prandtl number (Pr=$\frac{\rho c_p \upsilon}{\lambda}$) and Hartmann number ($\ Ha=B_0H\sqrt{\frac{\sigma}{\rho\vartheta}}$) in addition to Joule heating parameter($J=\frac{\sigma\ B_0^2u_{max}H}{\rho\ C_pT_{wall}}$) become apparent.\\
\FloatBarrier
\subsubsection{\textbf{Physics Informed Neural Network}}
\begin{figure}[h]
	\centering
	\includegraphics{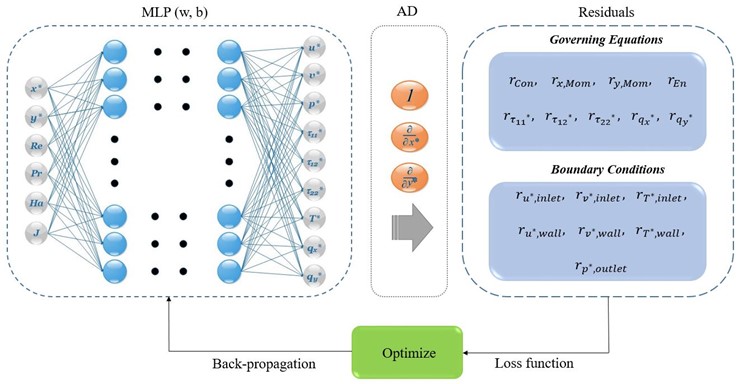}
	\caption{
	Schematics of the PINN used in the present study
	}
	\label{Fig2}
\end{figure}

\begin{figure}[h]
	\centering
	\includegraphics[scale=0.9]{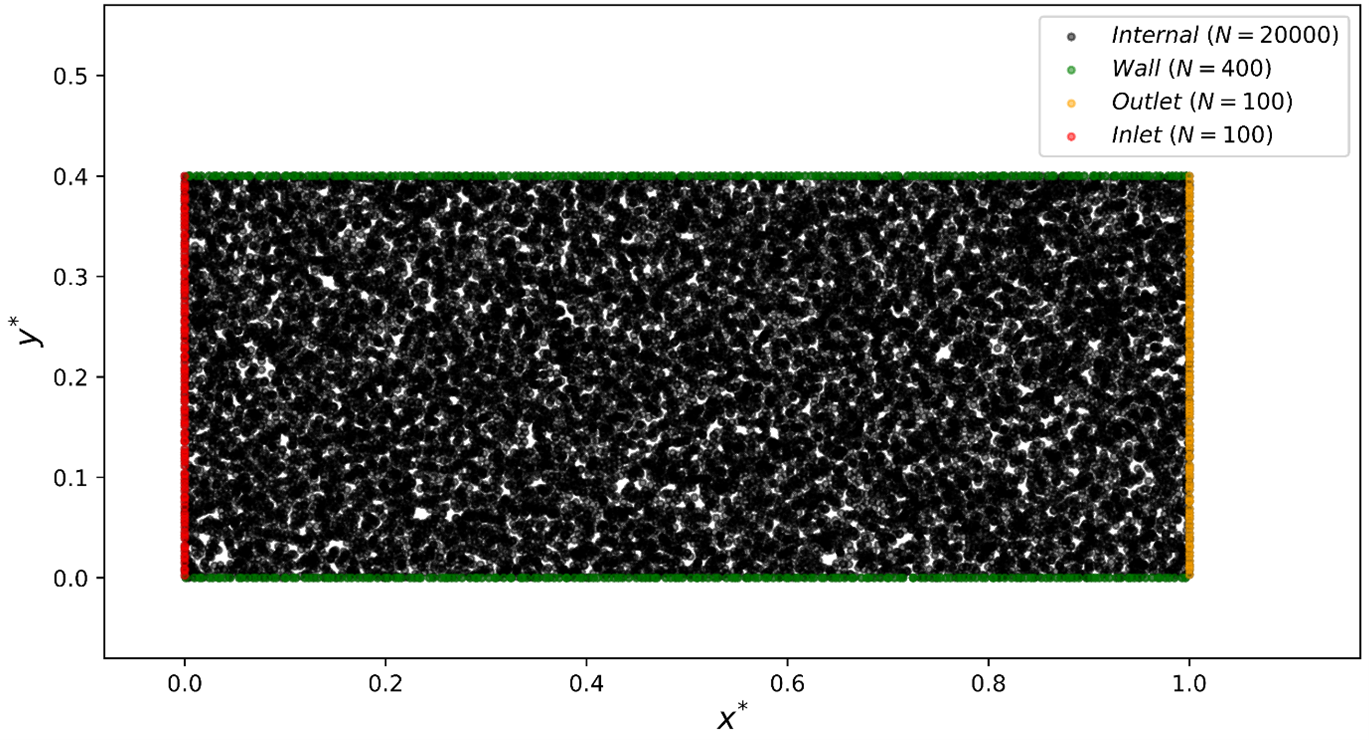}
	\caption{
points generated using the Latin Hypercube Sampling (LHS) method are used in the governing equations and boundary conditions in the 2-D domain.
	}
	\label{Fig3}
\end{figure}

In accordance with the previous section, several partial differential equations need to be considered in order to analyze the fluid flow in presence of magnetic fields. In order to overcome the shortcomings of the conventional methods as mentioned in the introduction, in this study the flow is analyzed using a method employing PINNs. This method is based on neural networks, in which the loss function is formed using governing equations and boundary conditions.
In this method employing MLP networks, first several physical and spatial dimensionless entities including $(J,Re,Pr,Ha)$ and $(y^\ast,x^\ast)$ are defined as the input parameters and quantities such as dimensionless velocity components ($u^\ast,v^\ast$), dimensionless pressure($p^\ast$) as well as dimensionless Cauchy stress tensor terms(${\tau_{22}}^\ast,{\tau_{12}}^\ast,{\tau_{11}}^\ast$), dimensionless temperature($T^\ast$) and dimensionless heat flux components ($q_x^\ast,q_y^\ast$) are defined as the output variables of the network. The network structure is depicted in Fig.\ref{Fig2}
\\
The derivative of the output variables is then calculated using Automatic Derivative (AD) method\cite{29} and substituted in the governing equation.
In order to incorporate the nonlinear and inherit complexity of the governing equations regarding the fluid flow in presence of magnetic fields, the $\tanh$ function is used as a nonlinear activation function in each neuron in the hidden layers of the network.
The loss function is then formed based on the physics governing the problem and boundary conditions. In this method, instead of discretization and mesh generation, several points within the domain and its boundary are randomly selected and the objective is to minimize the loss function of these points. In this study, {Latin Hypercube Sampling (LHS) algorithm} is employed to generate the collocation points as depicted in Fig.\ref{Fig3} \cite{30}. 20,000 points are considered in the 2-D domain in addition to the 600 points selected from the boundaries of the domain. The architecture of the said points is {portrayed} in Fig.\ref{Fig3}
\\
At the beginning, the weights of the neural networks are initialized using Xavier Initialization algorithm.\cite{31} updating of the network parameters such as biases and weights is carried out using optimization algorithms including ADAM\cite{32} and L-BFGS\cite{33}. Computations continue until the calculated overall loss is less than the expected value.
In order to define the loss function in this study, first the residuals are computed using Equations \ref{Eq24} to \ref{Eq39}. Equations \ref{Eq24} to \ref{Eq32} are related to the governing equations of the problem and Equations \ref{Eq33} to \ref{Eq39} pertain the imposed boundary conditions.
 In this set of equations, $r$ denotes the residual for each output variable. The loss function of the equations, boundary conditions (Dirichlet or Neumann) and {total loss functions} are defined by Eq.\ref{Eq40}, Eq.\ref{Eq41} and Eq.\ref{Eq42}
\FloatBarrier
\begin{align}
	&r_{Con}=p^\ast+\frac{{\tau_{11}}^\ast+{\tau_{22}}^\ast}{2}
	\label{Eq24}\\
	&r_{x,Mom}=u^\ast\frac{\partial\ u^\ast}{\partial\ x^\ast}+v^\ast\frac{\partial\ u^\ast}{\partial\ y^\ast}-\left(\frac{\partial{\tau_{11}}^\ast}{\partial\ x^\ast}+\frac{\partial{\tau_{12}}^\ast}{\partial\ y^\ast}\right)+\frac{{Ha}^2}{Re}u^\ast\\
	&r_{y,Mom}=u^\ast\frac{\partial\ v^\ast}{\partial\ x^\ast}+v^\ast\frac{\partial\ v^\ast}{\partial\ y^\ast}-(\frac{\partial{\tau_{12}}^\ast}{\partial\ x^\ast}+\frac{\partial{\tau_{22}}^\ast}{\partial\ y^\ast})\\
	&r_{En}=u^\ast\frac{\partial\ T^\ast}{\partial\ x^\ast}+v^\ast\frac{\partial\ T^\ast}{\partial\ y^\ast}-\frac{1}{RePr}\left(\frac{\partial{q_x}^\ast}{\partial\ x^\ast}+\frac{\partial{q_y}^\ast}{\partial\ y^\ast}\right)-J{u^\ast}^2\\
	&r_{{\tau_{11}}^\ast}=-p^\ast+\frac{2}{Re}\frac{\partial\ u^\ast}{\partial\ x^\ast}-{\tau_{11}}^\ast\\
	&r_{{\tau_{12}}^\ast}=-p^\ast+\frac{2}{Re}\frac{\partial\ v^\ast}{\partial\ y^\ast}-{\tau_{22}}^\ast\\
	&r_{{\tau_{22}}^\ast}=\frac{1}{Re}(\frac{\partial\ u^\ast}{\partial\ y^\ast}+\frac{\partial\ v^\ast}{\partial\ x^\ast})-{\tau_{12}}^\ast\\
	&r_{{q_x}^\ast}={q_x}^\ast-\frac{\partial\ T^\ast}{\partial\ x^\ast}\\
	&r_{{q_y}^\ast}={q_y}^\ast-\frac{\partial\ T^\ast}{\partial\ y^\ast}
	\label{Eq32}
	\\
	&r_{u^\ast,inlet}=u^\ast\mathrm{\mathrm{-}}\frac{u_{\mathrm{inlet}}}{\mathrm{u}_{\mathrm{max}}}
	\label{Eq33}
	\\
	&r_{v^\ast,inlet}=v^\ast\\
	&r_{T^\ast,inlet}=T^\ast\mathrm{\mathrm{-}}\frac{T_{\mathrm{inlet}}}{\mathrm{T}_{\mathrm{wall}}}\\
	&r_{u^\ast,wall}=u^\ast\\
	&r_{v^\ast,wall}=v^\ast\\
	&r_{T^\ast,wall}=T^\ast\mathrm{\mathrm{-}}\frac{T_{\mathrm{wall}}}{\mathrm{T}_{\mathrm{wall}}}\\
	&r_{p^\ast,outlet}=p^\ast-\frac{p_{outlet}}{\rho u_{\text{max}}^2}
	\label{Eq39}
\end{align}
\begin{equation}
	{Loss}_{PDE}=\frac{1}{N_{PDE}}\sum_{j=1}^{N_{PDE}}\left({r_{Con}}^2+{r_{x,Mom}}^2+{r_{y,Mom}}^2+{r_{En}}^2+{r_{{\tau_{11}}^\ast}}^2+{r_{{\tau_{12}}^\ast}}^2+{r_{{\tau_{22}}^\ast}}^2+{r_{{q_x}^\ast}}^2+{r_{{q_y}^\ast}}^2\right)
	\label{Eq40}
\end{equation}
\begin{dmath}
	{Loss}_{BC}=\frac{1}{N_{BC,inlet}}\sum_{j=1}^{N_{BC,inlet}}\left({r_{u^\ast,inlet}}^2+{r_{v^\ast,inlet}}^2+{r_{T^\ast,inlet}}^2\right)+\\
	\frac{1}{N_{BC,wall}}\sum_{j=1}^{N_{BC,wall}}\left({r_{u^\ast,wall}}^2+{r_{v^\ast,wall}}^2+{r_{T^\ast,wall}}^2\right)+\frac{1}{N_{BC,outlet}}\sum_{j=1}^{N_{BC,outlet}}\left({r_{p^\ast,outlet}}^2\right)
	\label{Eq41}
\end{dmath}
	\begin{dmath}
		{Loss}_{Total}={Loss}_{PDE}+{\beta\ Loss}_{BC}
		\label{Eq42}
	\end{dmath}
In this study, the loss function coefficient for boundary conditions ($\beta$) is assumed to be constants and set to one. The computations terminate when the {total} loss function is deemed lower than a specific threshold.
\FloatBarrier
\section{
Results \& Conclusion
}
\begin{figure}[h]
	\centering
	\includegraphics{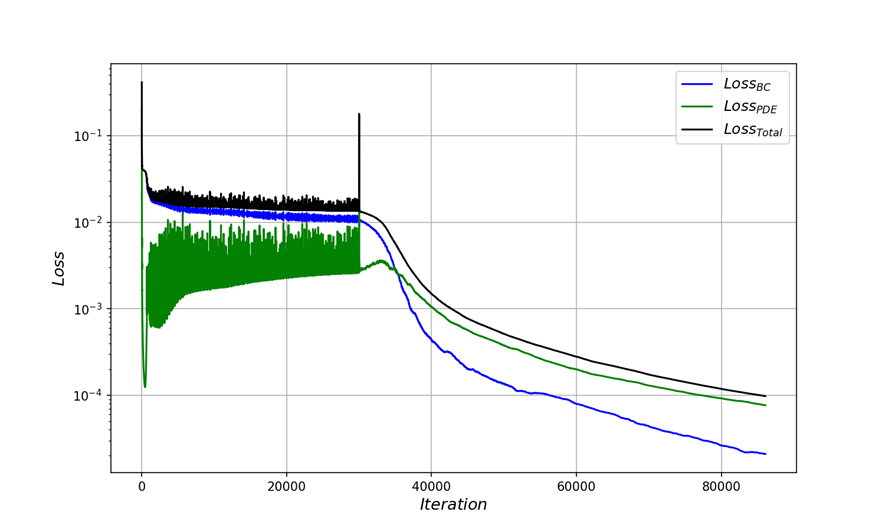}
	\caption{
	Loss functions in the Physics-Informed Neural Network plotted against iteration number
	}
	\label{Fig4}
\end{figure}

In the following section, the neural network described by Fig.\ref{Fig2} is applied to the problem at hand which concerns fluid mechanics and heat transfer under the influence of magnetic fields. This approach is implemented using Python\cite{34} environment and Pytorch library\cite{35}. Data regarding the independence of the obtained solution from the number of collocation points, layers and neurons in the network is included in the Appendix. The plot of loss function which includes loss functions of the governing equations, boundary conditions and {total} loss is depicted in Fig.\ref{Fig4}. Since the dimensionless forms of the governing equations are used, the {coefficients} of all terms in the loss function is assumed to be one. The convergence of the proposed method is evidenced by the decreasing trends of the loss functions, achieving errors below $10^{-4}$. The sudden increase observed in the 30000th iteration is attributed to the shift from ADAM optimization algorithm to L-BFGS algorithm. According to recommendations made by Rao et al,\cite{17} Eivazi et al,\cite{19} in addition to Biswass and Anald\cite{36}, the initial iterations are carried out using ADAM algorithm to avoid getting stuck in the local minimums and allow for the exploration of the problem mechanics. Also to achieve higher accuracy and convergence, L-BFGS method is employed in the later iterations.
\FloatBarrier
\subsection{\textbf{Investigating the dimensionless velocity components, Temperature and Pressure}}
\begin{figure}[h]
	\centering
	\includegraphics{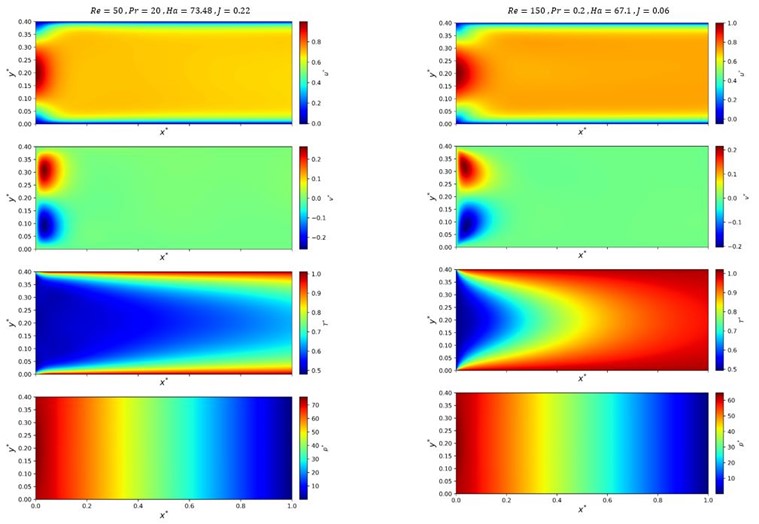}
	\caption{
	Contours of dimensionless velocity components, pressure and temperature obtained from parametric solutions using PINNs.
	}
	\label{Fig5}
\end{figure}

\begin{figure}
	\centering
	\includegraphics{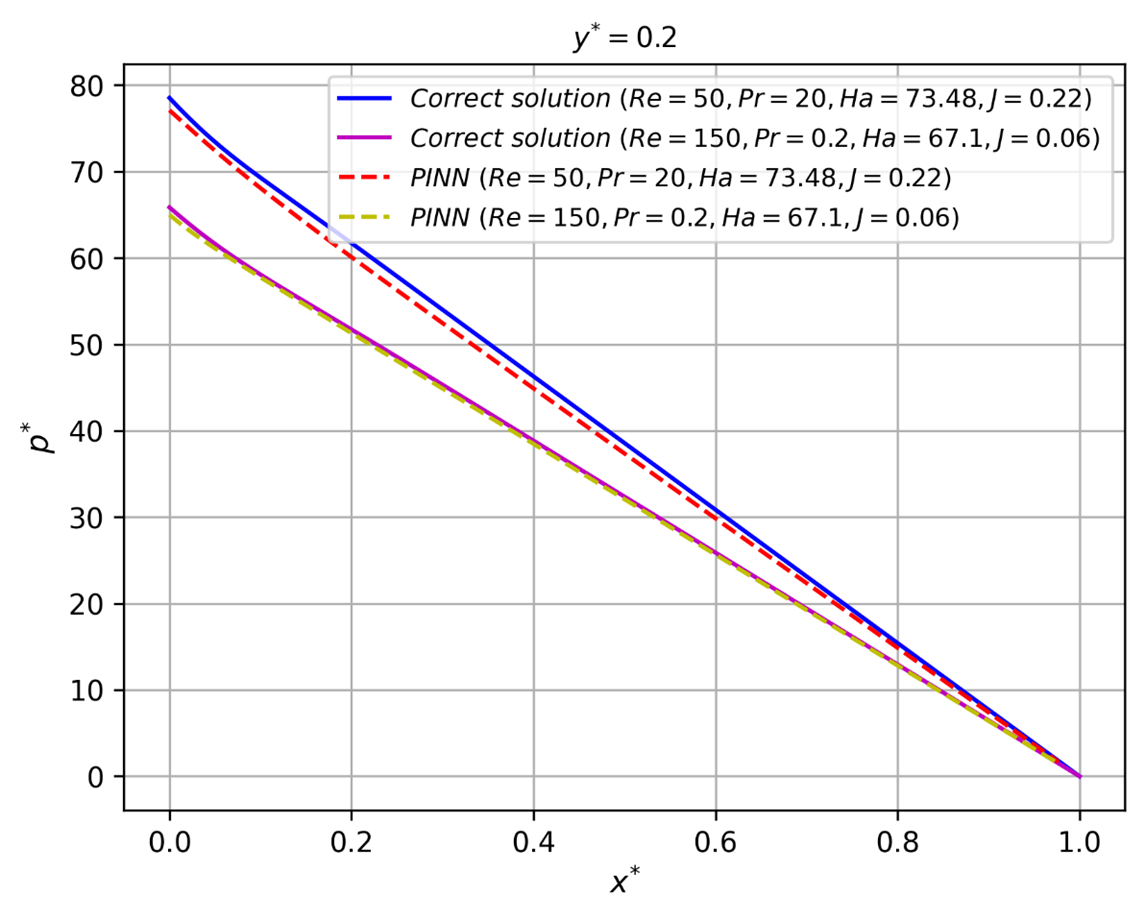}
	\caption{
	Trend of the dimensionless pressure along the centerline of the channel and comparison with the validated results.
	}
	\label{Fig6}
\end{figure}

\begin{figure}[h]
	\centering
	\includegraphics{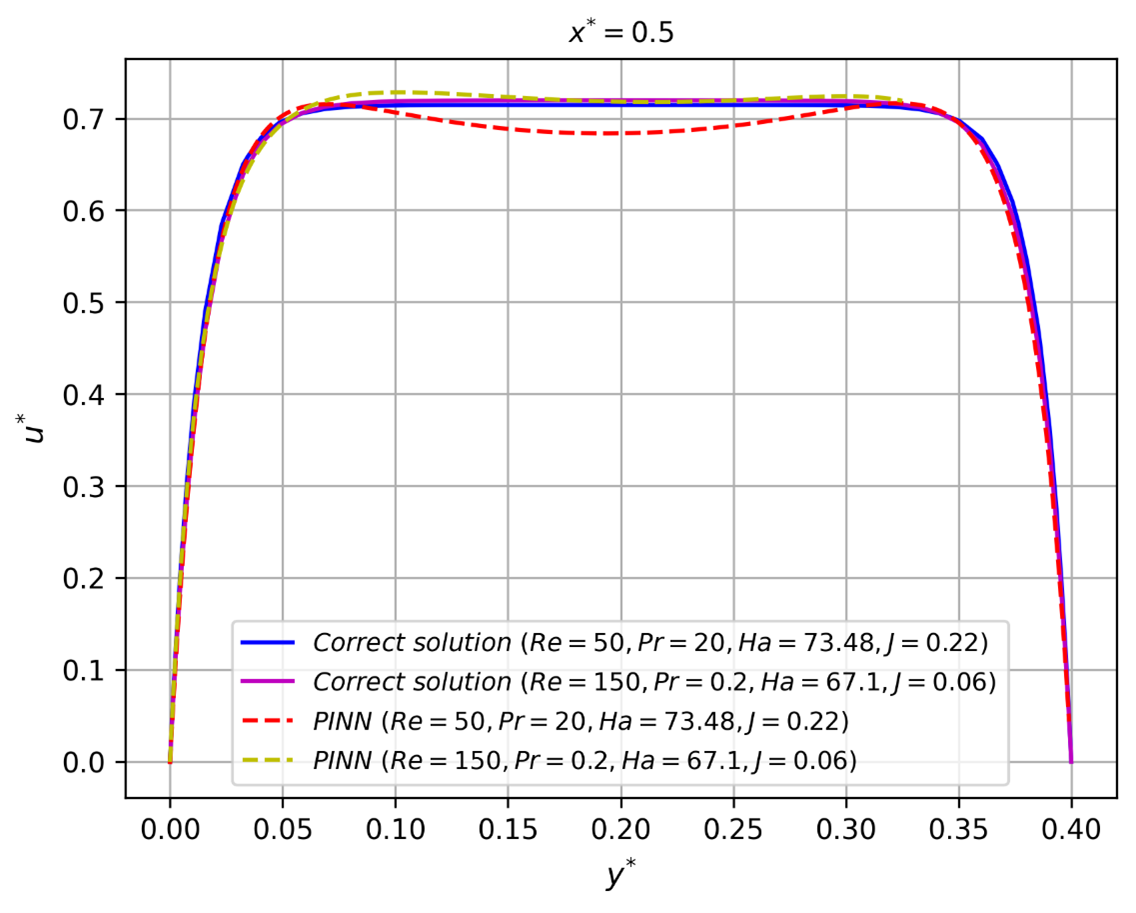}
	\caption{
	{distribution} of the horizontal component of the dimensionless velocity in the middle section of the channel and comparison with the available validated data
	}
	\label{Fig7}
\end{figure}

\begin{figure}[h]
	\centering
	\includegraphics{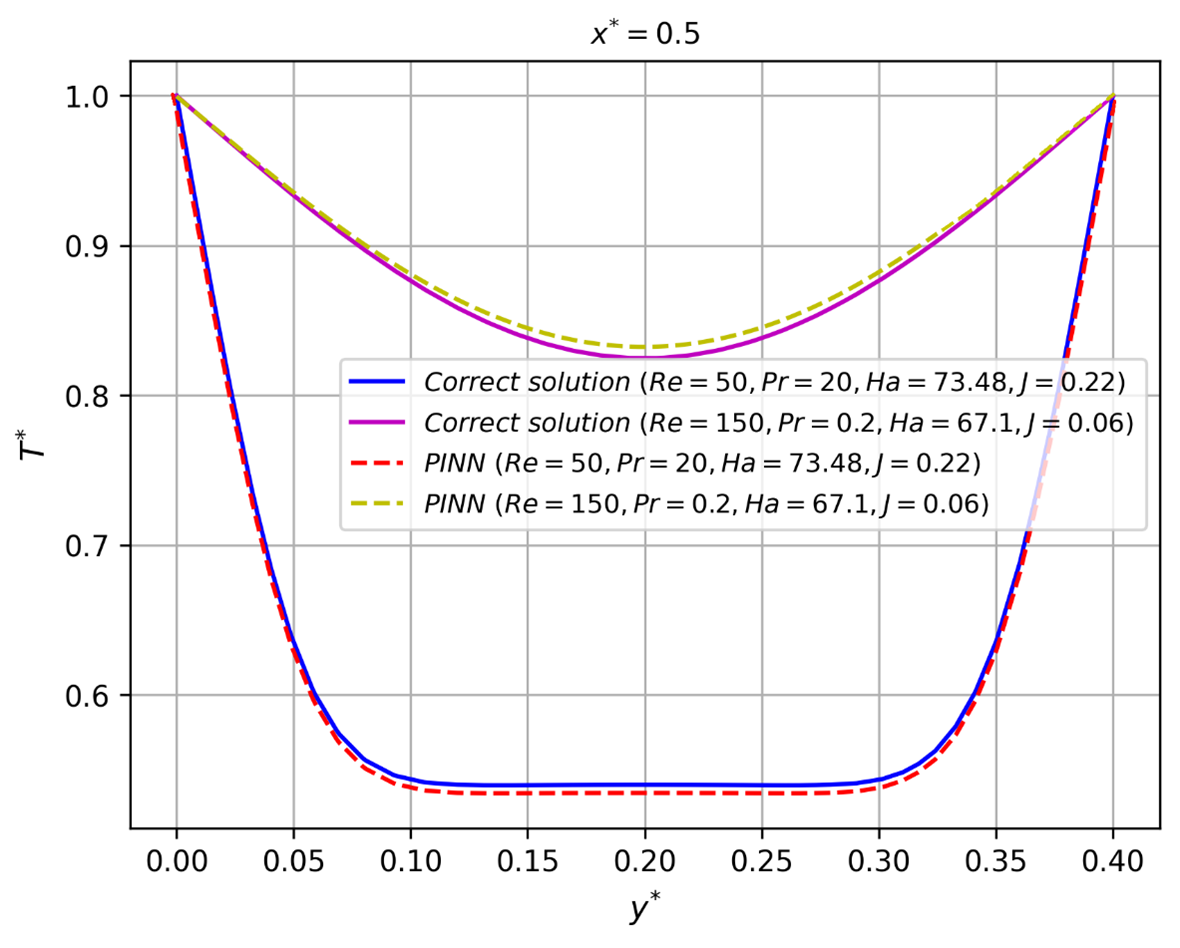}
	\caption{
	{variations in} the dimensionless temperature in the middle section of the channel and comparison with the validated results.
	}
	\label{Fig8}
\end{figure}

In this section, use of Physics-Informed Neural Networks in analyzing fluid flow and heat transfer under the influence of magnetic fields inside a channel is explored. Since the parametric solution to the problem has been obtained, physical parameters can be visualized in their respective range depicted in Table.\ref{Ranges}. Contours of dimensionless velocity components, pressure and temperature of two simulations whose physical characteristics are randomly selected are depicted in Fig.\ref{Fig5}
\begin{table}[h]
	\centering
	\begin{tabular}{|c|c|}
		\hline
		Parameter&Range\\
		\hline
		Re&5-200\\
		\hline
		Pr&0.01-100\\
		\hline
		Ha&5-200\\
		\hline
		J&0.001-1\\
		\hline
	\end{tabular}
	\label{Ranges}
	\caption{
	Ranges of Physical Parameters
	}
\end{table}
In order to validate the obtained results, the results have been put to test against data obtained from conventional computational methods. Furthermore, the trend of dimensionless pressure on the centerline of the channel obtained from the presented solution is captured within the range of parameters, As depicted in Fig.\ref{Fig6}. As the Heartmann number and Reynolds number increase, a rise in pressure drop becomes evident. Here, contribution of each of these parameters to the physics of the flow can be investigated by increasing one while decreasing another. The obtained results from PINN method conform well with the data obtained from CFD.
\FloatBarrier
Furthermore, as velocity contours at the middle of the channel obtained from the PINN method are compared against the data from the solution which employs CFD, adherence with the validated results becomes more evident as shown in Fig.\ref{Fig7}. It can be inferred from the results that as the Heartmann number increses, the velocity distribution becomes more uniform, making the plot of dimensionless velocity similar to that of the turbulent flow. 
The results concerning the dimensionless temperature in the middle section of the channel which are attained by the presented solution are compared against results obtained from CFD and their consistency with the validated results proves the accuracy of the proposed solution for cases involving fluid flow and heat transfer as depicted in Fig.\ref{Fig8}. As the Heartmann number and Joule heating parameter increase, the heat transfer of the channel is perceived to rise as well. Additionally, as the Reynolds and Prandtl numbers increase, the temperature of the fluid is seen to rise as well as channel heat transfer. The alignment of the obtained results from the proposed method with those obtained from other computational solutions prove its profound ability to solve the equations governing the fluid behavior. 
\FloatBarrier
\subsection{\textbf{Effect of Joule heating parameter}}
\begin{figure}[h]
	\centering
	\includegraphics{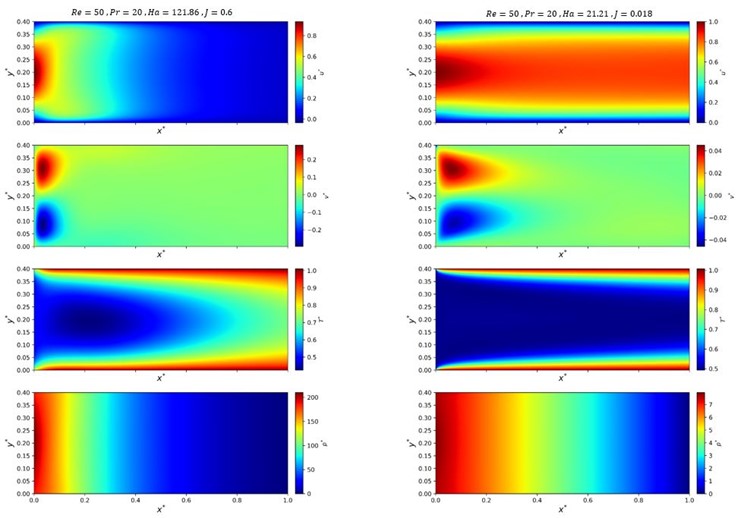}
	\caption{
		dimensionless velocity components, pressure and temperature for the two distinct random values of Joule heating parameter from the upper and lower range.
	}
	\label{Fig9}
\end{figure}

\begin{figure}[h]
	\centering
	\includegraphics{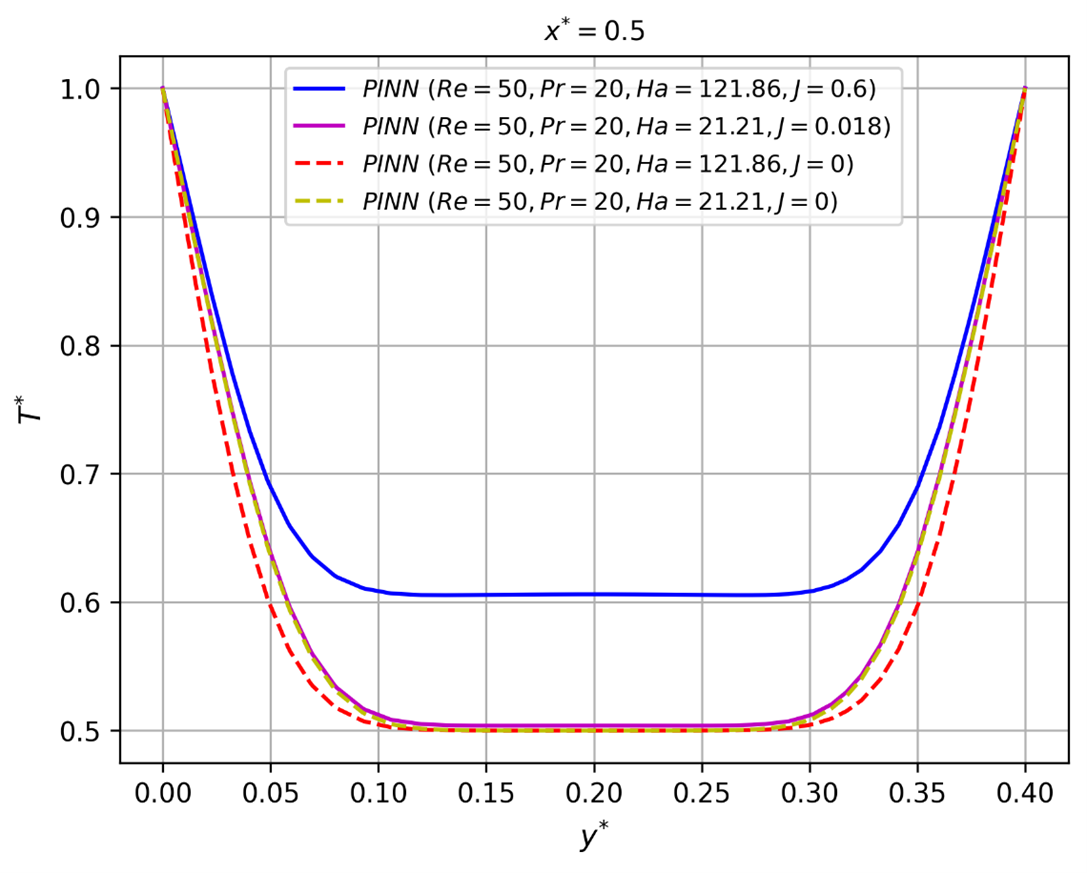}
	\caption{
	Comparison between the dimensionless temperature profile in the middle section of the channel for two cases where the Joule heating parameters was considered or not.
	}
	\label{Fig10}
\end{figure}

As the magnetic fields are applied to the fluid which is also {electrically conductive}, a source term appears in the momentum equations. Also, the presence of magnetic field also manifests itself as a heat source term in the energy equation. This section explores the importance and consequences of incorporating the magnetic fields in the energy equation. In Fig.\ref{Fig9} the dimensionless velocity component contours, dimensionless pressure and dimensionless temperature are depicted for two distinct random set of values selected from the lower and higher range of parameters related to the magnetic field.
 To conduct more comprehensive investigations, in Fig.\ref{Fig10} the contour of dimensionless temperature in the middle section of the channel ($x^\ast=\frac{1}{2}$) is illustrated for cases were the Joule heating parameter was incorporated or disregarded, having a lesser value or having a substantial value. As shown, as the electrical characteristics of the flow become more pronounced or stronger magnetic fields are taken into consideration, the influence of the Joule heating parameter becomes more considerable. In other words, the effects of magnetic fields should also be incorporated in the energy equation in addition to being implemented in the momentum equations. 
\FloatBarrier
\subsection{\textbf{Neumann Boundary Condition (Constant heat flux on the wall)}}
To further assess the ability of the proposed method to tackle problems with Neumann boundary conditions, the dimensionless velocity, pressure and temperature distributions are depicted in  Fig\ref{Fig11} where it is assumed that the upper boundary of the channel is subjected to constant heat flux while the bottom wall is assumed to be insulated. The dimensionless temperature in the middle section of the channel as well as trend of pressure along the channel are compared with the data obtained from conventional solutions. {The results indicate that the method at hand is capable of solving problems with Neumann boundary condition.
\begin{figure}[h]
	\centering
	\includegraphics[scale=0.75]{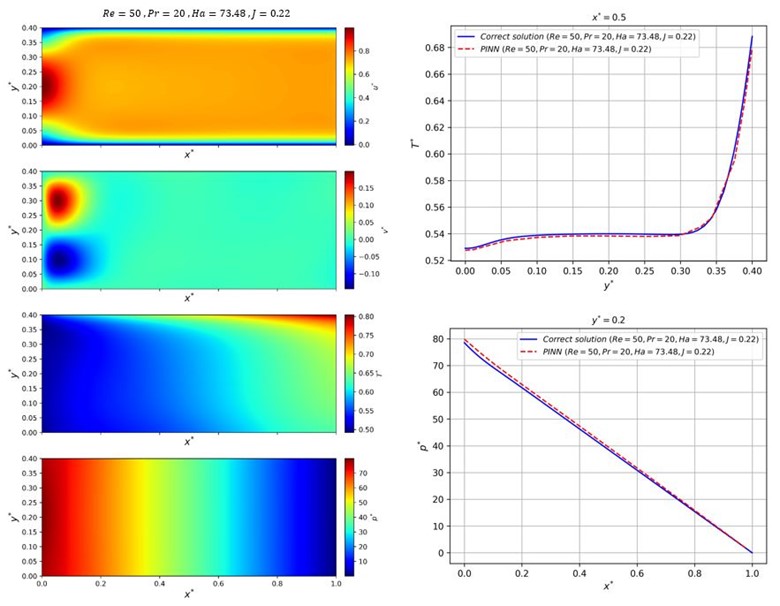}
	\caption{comparing the results obtained from the described approach with the validated results.}
	\label{Fig11}
\end{figure}
\FloatBarrier
\subsection{\textbf{Extrapolation of the problem for parameters outside of the predefined range}}
To further assess the ability of the proposed method to tackle problems where the input parameters fall outside of the predefined range, several intervals have been considered as illustrated in Fig.\ref{Fig12}. Several problems are then solved under different configurations defined by dimensionless velocity components, pressure and temperature. Comparing the changes in the dimensionless pressure profile along the centerline of the channel with the data obtained through conventional CFD methods indicates that the said method can also deduce the flow characteristics outside of the parameter predefined ranges, namely the Unseen values. Furthermore, considering the fact that the physical parameters have acquired values outside of the said ranges, a considerable deviation from the correct results obtained from CFD is perceived. In general, fluid flow and heat transfer inside the channel under the influence of magnetic fields have been modeled with reasonable accuracy, indicating generalization ability of the method at hand.
\begin{figure}[h]
	\centering
	\includegraphics{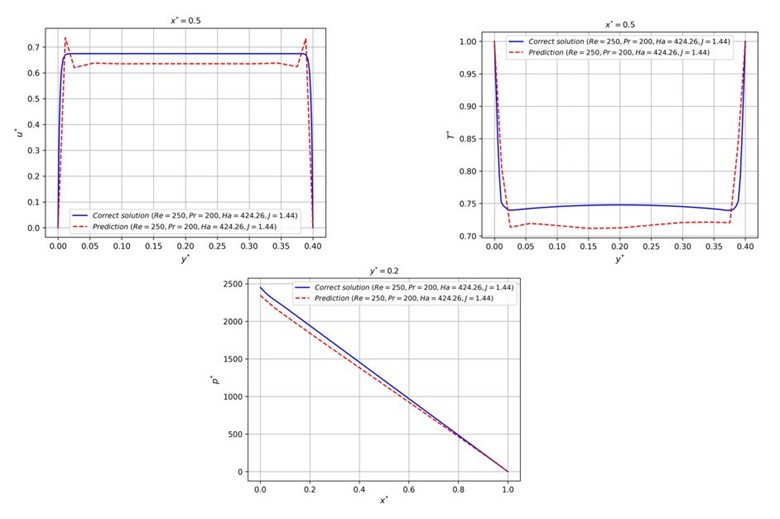}
	\caption{
	Comparing the results obtained from PINN method with validated data
	}
	\label{Fig12}
\end{figure}
\FloatBarrier
\subsection{
\textbf{Entropy Generation}
}
Entropy generation is often used as a tool to measure the depletion of the available energy in form of a metric to better monitor irreversibilities within a process. In other words, in thermodynamic systems as less entropy is generated, the efficiency rises since less energy is wasted. Local entropy generation, disregarding entropy generation associated with {viscous dissipation}, is defined according to the following equation and such definition is adopted throughout this article\cite{41}-\cite{44} 
\begin{equation}
	S_{gen}=\frac{1}{\lambda\ T^2}(q_x^2+q_y^2)+\frac{\sigma\ B_0^2}{T}u^2
\end{equation}
Since all physical parameters are represented by a dimensionless entity, entropy generation is then defined as follows:
\begin{equation}
	{S_{gen}}^\ast=\frac{S_{gen}H^2}{\lambda}=\frac{1}{{T^\ast}^2}\left({{q_x}^\ast}^2+{{q_y}^\ast}^2\right)+\frac{RePrJ}{T^\ast}{u^\ast}^2
\end{equation}
{The subsequent right-hand side} of the equation represents the effect of Joule heating parameter defined as ${S_{gen,J}}^\ast=\frac{RePrJ}{T^\ast}{u^\ast}^2$. In Fig.\ref{Fig13} the ratio of entropy generation associated with Joule heating parameter to the {overall} entropy generation in the channel is visualized. Because of the significant temperature gradient near the walls of the channel, the entropy generation as a result of heat transfer acquires high values within this region. Also the Joule heating parameter contributes to the entropy generation at the centerline of the channel. As the effects of magnetic field become more pronounced in form of higher values of Joule heating parameter, the entropy generation becomes more considerable, underscoring the importance of including Joule heating parameter in calculation of entropy in presence of magnetic fields.
\begin{figure}[h]
	\centering
	\includegraphics{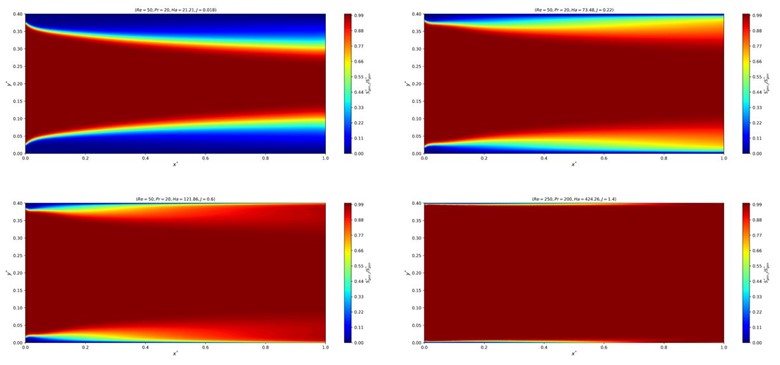}
	\caption{Evaluating the contribution of the Joule heating parameter in entropy generation in form of $\frac{{S_{gen,J}}^\ast}{{S_{gen}}^\ast}$ using a parametric study employing PINN method}
	\label{Fig13}
\end{figure} 
A comparison between the results acquired from PINN method and conventional CFD method has been carried out to assure the validity of the results. Variations in entropy generation associated with Joule heating is calculated using parametric PINN method as depicted in Fig.\ref{Fig14}. As the Joule heating increases, which is associated with the enhanced impact of magnetic fields, there is a corresponding rise in entropy generation within the channel. This leads to a more uniform profile, showcasing a widespread increase throughout the channel's entire section. It is also noteworthy that in cases where the parameters fell outside of the predefined ranges, the proposed method was able to attain reasonable results and it is thus evident that the acquired results conform well with the data obtained from conventional computational method. 
\begin{figure}[h]
	\centering
	\includegraphics{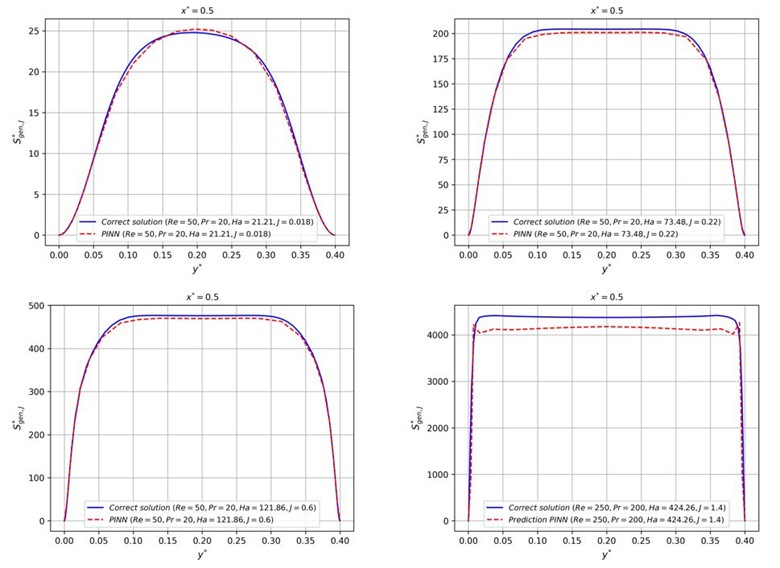}
	\caption{
	changes in the entropy generation associated with Joule heating in the middle section of the channel obtained from the proposed method and conventional methods 
	}
	\label{Fig14}
\end{figure}
\FloatBarrier
\subsection{\textbf{The Ill-posed problem}}
At last, the ill-posed problem is explored using the PINN method with the geometry and collocating points according to Fig.\ref{Fig3}. Solving this category of the problems using the conventional computational methods is difficult but the flexibility of the PINN method allows exploring complicated ill-posed problems. An instance of this category would be problems which involve calculating the Joule heating parameter as an unknown. As new unknowns are introduced in the problem, several parameters such as velocity components and pressure or temperature of several points within the domain have to be specified. In this study, it is assumed that in 
context
 concerning heat transfer in the channel the goal is to find the heat source term associated with the magnetic field in order to achieve a certain temperature at the outlet of the {channel}. Consequently, the temperature at the outlet of the channel is used to define a new loss term of the output variables of the neural network to enhance the learning process needed to obtain the Joule heating parameter as shown in Fig.\ref{Fig13}.
\begin{figure}
	\centering
	\includegraphics{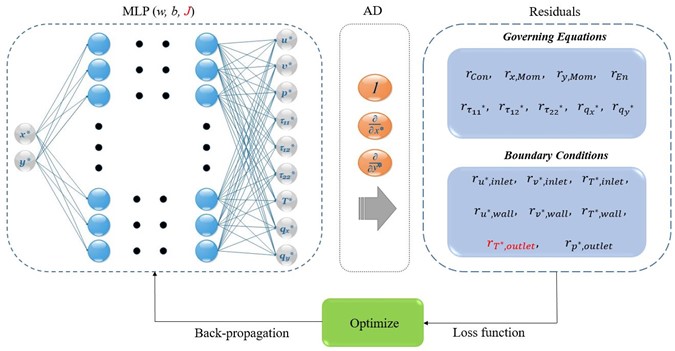}
	\caption{
	Schematics of the neural network used for solving the {ill-posed} problem
	}
	\label{Fig15}
\end{figure}
The residual of the outlet temperature present in Fig.\ref{Fig13} is defined using Eq.\ref{Eq43}.
\begin{equation}
	r_{T^\ast,outlet}=T^\ast\mathrm{\mathrm{-}}\frac{T_{\mathrm{outlet}}}{\mathrm{T}_{\mathrm{wall}}}
	\label{Eq43}
\end{equation}
Moreover, the Eq.\ref{Eq41} has been modified and Eq.\ref{Eq44} has been obtained.
\begin{dmath}
	{Loss}_{BC}=\frac{1}{N_{BC,inlet}}\sum_{j=1}^{N_{BC,inlet}}\left({r_{u^\ast,inlet}}^2+{r_{v^\ast,inlet}}^2+{r_{T^\ast,inlet}}^2\right)+\frac{1}{N_{BC,wall}}\sum_{j=1}^{N_{BC,wall}}\left({r_{u^\ast,wall}}^2+\\{r_{v^\ast,wall}}^2+{r_{T^\ast,wall}}^2\right)+\frac{1}{N_{BC,outlet}}\sum_{j=1}^{N_{BC,outlet}}\left({r_{p^\ast,outlet}}^2+{r_{T^\ast,outlet}}^2\right)
	\label{Eq44}
\end{dmath}
In order to solve this problem using PINN method, the Joule Heating parameter has to be optimized alongside weights and biases of the neural network. Similar to the weights and biases of the network, a random value has to be designated to the Joule heating parameter. If a bad initial guess is made for this number at the beginning, obtaining the solution becomes more difficult and more iterations are needed to achieve the desired value. In Fig.\ref{Fig14} the steps needed to be taken in order to calculate the Joule heating parameter using PINN are depicted. The convergence is achieved in 75000th iteration.
\begin{figure}
	\centering
	\includegraphics{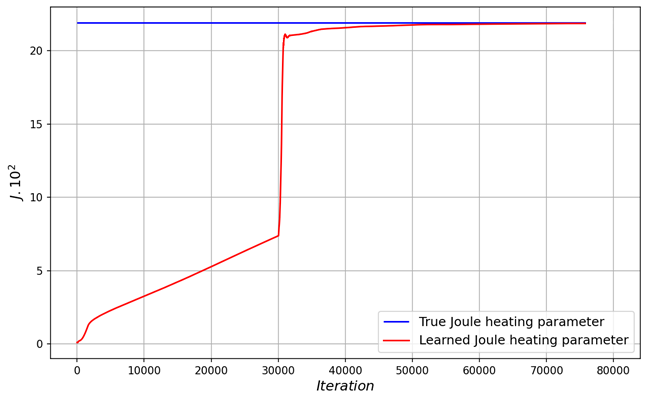}
	\caption{
	Obtaining the Joule heating parameter using PINNs
	}
	\label{Fig16}
\end{figure}
The results obtained in this study depict the profound ability of the proposed method to model heat transfer and fluid {flow} in the channel under the influence of magnetic fields. One of the main advantages of the proposed method in this research is not needing labeled data for the input of the neural network. {Acquiring labeled data is impractical in many situations and not feasible in other contexts.}\cite{37}. This method has proved to be able to tackle ill-posed problems and obtain parametric solutions for problems often encountered in context of engineering. It has been shown that combining this method with the conventional method can enhance their performance\cite{38}. More studies can be conducted for selecting the parameters of the neural network and its structure in more complex problems\cite{39}. The present study has also been used to model fluid flow in channels having more complex geometries where the geometric parameters have been included and accurate results have been achieved.\cite{45}
\FloatBarrier
\subsection{\textbf{Conclusion}}
In this study, heat transfer and entropy generation in a channel subjected to magnetic fields is investigated. Governing equations have been modified to incorporate low-order derivatives and the dimensionless forms of the equations are implemented in calculations. Using the low-order derivatives in the equations reduces the calculation time and helps with implementation of the Neumann boundary condition as well as facilitating steps required for calculating entropy generation. Using dimensionless numbers reduces the parameters involved and creates a balance between the loss function terms. Convergence is also enhanced when dimensionless parameters are employed. Dimensionless numbers governing the physics of the problem are involved in the learning process as inputs to the network and the impact of each parameter is also studied in this research. Furthermore, role of the Joule heating parameter as a heat source term in the energy equation and a vital parameter in the calculations regarding entropy generation is investigated. In order to validate the proposed method when applied to different boundary conditions, a boundary condition defined on the basis of heat flux is also taken into consideration. In another section of this research, solution to the ill-posed problem and obtaining the Joule heating parameter using the PINN method is explored. The results indicate that the PINN method is capable of calculating this dimensionless number with great precision. In this study, a comprehensive study is conducted, exploring the physical parameters of the fluid flow, heat transfer and entropy generation in the channel subjected to magnetic fields.\\\FloatBarrier
\vspace{1cm}
\noindent\textbf{\Large{CRediT authorship contribution statement}}\\
\textbf{\small{Ehsan Ghaderi}}: Conceptualization, Methodology, Visualization, Software, Writing the original draft.\\
\textbf{\small{Bijarchi et al.}}: Conceptualization, Review and Editing, Supervision.\\\\
\textbf{\Large{Acknowledgment}}\\
We would like to thank Nvidia for providing the GPUs for this work.
\\\\\textbf{\Large{Conflict of Interest}}\\
The authors declare no conflict of interest\\
\\\textbf{\Large{Data Availability}}\\
The python codes used to generate the results presented are available on Github.
\\\url{https://github.com/ehsangh94/PINN}

\pagebreak
\FloatBarrier
\section*{Appendix}
{In order to solve the problem using this method, several points must be selected from the domain which need to satisfy the governing equations.} In contrast to the conventional methods, instead of domain discretization and grid generation, a set of points are randomly selected within the domain using the LHS algorithm.
optimal number of points needed for calculations is a parameter of interest in this article.
While modeling complex physical phenomenons, more points need to be selected in regions near boundaries where drastic changes are expected. 
To examine the independence of the model from the collocating points, several simulations are carried out and their results are tabulated in Table.\ref{Nums}. It is noteworthy that since the problem at hand is six dimensional, more points have to be selected relative to the 2-D problem. 
\begin{table}[h]
	\centering
	\caption{Effect of number of points in PINN method}
	\begin{tabular}{|c|c|c|c|}
		\hline
		\begin{tabular}{@{}c@{}}Loss function of \\ PINN method\end{tabular}&Computation time&\begin{tabular}{@{}c@{}}Number of \\ Points on boundaries\end{tabular}& Number of points in the domain\\
		\hline
		$2\times10^{-4}$&20min. 5sec&300&10000\\
		\hline
		$9\times10^{-5}$&23min. 42sec&600&20000\\
		\hline
		$1\times 10^{-4}$&63min. 4sec&2400&80000\\
		\hline
		$8\times10^{-3}$&101min. 23sec&4800&160000\\
		\hline
	\end{tabular}
	\label{Nums}
\end{table}
In Table.\ref{Nums} the impact of number of points on loss functions and computation time is shown. As more points are used in the simulations, error does not necessarily decrease while computation times increases substantially. Evidently, using more points in the simulation does not decrease the loss function since the algorithm is not able to decrease the loss function effectively. 
Proper number of hidden layers and number of neurons in each layer is also another important parameter investigated in this paper.
To find the most {efficient} architecture meeting accuracy requirements, several instances of different architectures {ranging from a single layer composed of ten neurons to nine layers each consisting of 100 neurons are investigated.}
Error for temperature distribution in different architectures is examined and In general, it can be inferred that the errors are reduced when the network becomes deeper(meaning to have more layers) and wider(meaning to have more neurons in each layer).
The optimal architecture of the network is then selected such that no appreciable change in error is perceived when the size of the network increases.\cite{26} 
In Table.\ref{Nums2}, the effect of number of layers and number of neurons in each layer is perceivable. As deeper neural networks are used, the network exhibits more nonlinear behavior, leading to more precise results.
More complicated network means more computation time and will not yield a significantly more accurate result. For example, when the number of hidden layers is increased to seven and the neurons to fifty, the loss exhibits minuscule changes compared to simpler network.
These findings point out that substantial increase of the number of neurons and layers will cause the network to become too complex while accuracy remains mostly the same, as the computation time grows significantly.
\begin{table}[h]
	\centering
	\caption{
		Effect of number of hidden layers and neurons in each layer in PINN method
	}
	\begin{tabular}{|c|c|c|c|}
		\hline
		\begin{tabular}{@{}c@{}}Loss function of \\ PINN method\end{tabular}&Computation time&Number of neurons in each hidden layer&Number of hidden layers\\
		\hline
		$4\times10^{-2}$&15min. 8sec&10&1\\
		\hline
		$9\times10^{-3}$&28min. 5sec&20&3\\
		\hline
		$1\times10^{-4}$&49min. 31sec&30&5\\
		\hline
		$8\times10^{-5}$&70min. 2sec&50&7\\
		\hline
		$3\times10^{-5}$&114min. 34sec&100&9\\
		\hline
	\end{tabular}
	\label{Nums2}
\end{table}\\
According to this analysis, the network utilized for this research has five hidden layers, each having 30 neurons. A total of 41200 points are also deemed sufficient.\newline

\begin{thebibliography}{100}
	\bibitem{1}Shi S, Niu J, Wu Z, Luo S, Gao X, Fang Y, Zhang Z. Experimental and numerical investigation on heat transfer enhancement of vertical triplex tube heat exchanger with fractal fins for latent thermal energy storage. International Journal of Heat and Mass Transfer. 2022.
	\bibitem{2}Rasheed AH, Alias HB, Salman SD. Experimental and numerical investigations of heat transfer enhancement in shell and helically microtube heat exchanger using nanofluids. International Journal of Thermal Sciences. 2021.
	\bibitem{3}Wang YB, Huang LF, Lan N, Wang SL, Zhang BX, Yang YR, Wang XD, Lee DJ. Heat transfer enhancement by electrohydrodynamics in wavy channels. Applied Thermal Engineering. 2024.
	\bibitem{4}Fico F, Langella I, Xia H. Large-eddy simulation of magnetohydrodynamics and heat transfer in annular pipe liquid metal flow. Physics of Fluids. 2023.
	\bibitem{5}Yang X, Yu J, Guo Z, Jin L, He YL. Role of porous metal foam on the heat transfer enhancement for a thermal energy storage tube. Applied Energy. 2019.
	\bibitem{6}Mousa MH, Miljkovic N, Nawaz K. Review of heat transfer enhancement techniques for single phase flows. Renewable and Sustainable Energy Reviews. 2021.
	\bibitem{7}Dulikravich GS, Lynn SR. Unified electro-magneto-fluid dynamics (EMFD): A survey of mathematical models. International Journal of Non-Linear Mechanics. 1997 Sep 1;32(5):923-32.
	\bibitem{8}Davidson PA. Introduction to Magnetohydrodynamics. Cambridge University Press; 2016 Dec 22.
	\bibitem{9}Sheikholeslami M, Rokni HB. Simulation of nanofluid heat transfer in presence of magnetic field: A review. International Journal of Heat and Mass Transfer. 2017 Dec 1;115:1203-33.
	\bibitem{10}Mousavi SM, Ehteshami B, Darzi AA. Two-and-three-dimensional analysis of Joule and viscous heating effects on MHD nanofluid forced convection in microchannels. Thermal Science and Engineering Progress. 2021 Oct 1;25:100983.
	\bibitem{11}Anderson, D., Tannehill, J.C., Pletcher, R.H., Munipalli, R., Shankar, V.: Computational Fluid Mechanics and Heat Transfer. CRC press, Boca Raton, Florida, USA (2020)
	\bibitem{12}I. Goodfellow, Y. Bengio, and A. Courville, Deep learning. MIT press, 2016.
	\bibitem{13}LeCun Y, Bengio Y, Hinton G. Deep learning. nature. 2015 May 28;521(7553):436-44.
	\bibitem{14}Nathan Baker, Frank Alexander, Timo Bremer, Aric Hagberg, Yannis Kevrekidis, Habib Najm, Manish Parashar, Abani Patra, James Sethian, Stefan Wild, et al. Workshop report on basic research needs for scientific machine learning: Core technologies for artificial intelligence. Technical report, USDOE Office of Science (SC), Washington, DC (United States), 2019.
	\bibitem{15}M. Raissi, P. Perdikaris, and G. E. Karniadakis, “Physics-informed neural networks: A deep learning framework for solving forward and inverse problems involving nonlinear partial differential equations,” J. Comput. Phys., vol. 378, pp. 686–707, 2019.
	\bibitem{16}Raissi M, Yazdani A, Karniadakis GE. Hidden fluid mechanics: Learning velocity and pressure fields from flow visualizations. Science. 2020 Feb 28;367(6481):1026-30.
	\bibitem{17}C. Rao, H. Sun, and Y. Liu, “Physics-informed deep learning for incompressible laminar flows,” Theor. Appl. Mech. Lett., vol. 10, no. 3, pp. 207–212, 2020.
	\bibitem{18}A. Arzani, J. Wang, R. M. D. Souza, and R. M. D’Souza, “Uncovering near-wall blood flow from sparse data with physics-informed neural networks,” Phys. Fluids, vol. 33, no. 7, p. 71905, 2021.
	\bibitem{19}H. Eivazi, M. Tahani, P. Schlatter, and R. Vinuesa, “Physics-informed neural networks for solving Reynolds-averaged Navier–Stokes equations,” Phys. Fluids, vol. 34, no. 7, p. 75117, 2022.
	\bibitem{20}S. Cai, Z. Wang, S. Wang, P. Perdikaris, and G. E. Karniadakis, “Physics-informed neural networks for heat transfer problems,” J. Heat Transfer, vol. 143, no. 6, 2021.
	\bibitem{21}X. Meng, Z. Li, G. E. Karniadakis, D. Zhang, and G. E. Karniadakis, “PPINN: Parareal physics-informed neural network for time-dependent PDEs,” Comput. Methods Appl. Mech. Eng., vol. 370, no. September, p. 113250, 2020.
	\bibitem{22}H. Wang, R. Planas, A. Chandramowlishwaran, and R. Bostanabad, “Mosaic flows: A transferable deep learning framework for solving PDEs on unseen domains,” Comput. Methods Appl. Mech. Eng., vol. 389, p. 114424, 2022.
	\bibitem{23}G. E. Karniadakis, G. Street, A. D. Jagtap, and G. E. Karniadakis, “Extended Physics-informed Neural Networks (XPINNs): A Generalized Space-Time Domain Decomposition based Deep Learning Framework for Nonlinear Partial Differential Equations.,” in AAAI Spring Symposium: MLPS, 2021, no. 11.
	\bibitem{24}Barletta A, Celli M. Mixed convection MHD flow in a vertical channel: effects of Joule heating and viscous dissipation. International journal of heat and mass transfer. 2008 Dec 1;51(25-26):6110-7.
	\bibitem{25}Hossain MA. Viscous and Joule heating effects on MHD free convection flow with variable plate temperature. International Centre for Theoretical Physics; 1990.
	\bibitem{26}Laubscher R. Simulation of multi-species flow and heat transfer using physics-informed neural networks. Physics of Fluids. 2021 Aug 1;33(8).
	\bibitem{27}Hu C, Cui Y, Zhang W, Qian F, Wang H, Wang Q, Zhao C. Solution of conservative-form transport equations with physics-informed neural network. International Journal of Heat and Mass Transfer. 2023 Dec 1;216:124546.
	\bibitem{28}L. Bergman Theodore, s Lavine Adrienne, P. Incropera Frank, P. Dewitt David, Fundamentals of Heat and Mass Transfer, eighth ed., John Wiley \& Sons, US, 2018.
	\bibitem{29}A. G. Baydin, B. A. Pearlmutter, A. A. Radul, and J. M. Siskind, “Automatic differentiation in machine learning: a survey,” J. Marchine Learn. Res., vol. 18, pp. 1–43, 2018.
	\bibitem{30}Stein M. Large sample properties of simulations using Latin hypercube sampling. Technometrics. 1987 May 1;29(2):143-51.
	\bibitem{31}S. K. Kumar, “On weight initialization in deep neural networks,” arXiv Prepr. arXiv1704.08863, pp. 1–9, 2017.
	\bibitem{32}D. P. Kingma and J. L. Ba, “Adam: A method for stochastic optimization,” arXiv Prepr. arXiv1412.6980, pp. 1–15, 2014.
	\bibitem{33}D. C. Liu, and J. Nocedal, “On the limited memory BFGS method for large scale optimization,” Math. Program., vol. 45, no. 1, pp. 503–528, 1989.
	\bibitem{34}F. F. Chollet, Deep learning with Python. Simon and Schuster, 2021.
	\bibitem{35}Paszke, Adam, Sam Gross, Francisco Massa, Adam Lerer, James Bradbury, Gregory Chanan, Trevor Killeen et al. "Pytorch: An imperative style, high-performance deep learning library." Advances in neural information processing systems 32, 2019.
	\bibitem{36}Biswas SK, Anand NK. Three-dimensional laminar flow using physics informed deep neural networks. Physics of Fluids. 2023 Dec 1;35(12).
	\bibitem{37}L. Sun, H. Gao, S. Pan, and J. Wang, “Surrogate modeling for fluid flows based on physics-constrained deep learning without simulation data,” Comput. Methods Appl. Mech. Eng., vol. 361, p. 112732, 2020.
	\bibitem{38}M. Aliakbari, M. Mahmoudi, A. Arzani, P. Vadasz, and A. Arzani, “Predicting high-fidelity multiphysics data from low-fidelity fluid flow and transport solvers using physics-informed neural networks,” Int. J. Heat Fluid Flow, vol. 96, no. May, p. 109002, 2022.
	\bibitem{39}Zhao X, Gong Z, Zhang Y, Yao W, Chen X. Physics-informed convolutional neural networks for temperature field prediction of heat source layout without labeled data. Engineering Applications of Artificial Intelligence. 2023 Jan 1;117:105516.
	\bibitem{40}Wang, S., Nikfar, M., Agar, J.C. and Liu, Y., Stacked Deep Learning Models for Fast Approximations of Steady-State Navier-Stokes Equations for Low Re Flow. Intelligent Computing.
	\bibitem{41}Mahian O, Oztop H, Pop I, Mahmud S, Wongwises S. Entropy generation between two vertical cylinders in the presence of MHD flow subjected to constant wall temperature. International communications in heat and mass transfer. 2013 May 1;44:87-92.
	\bibitem{42}Ibáñez G, López A, Pantoja J, Moreira J. Entropy generation analysis of a nanofluid flow in MHD porous microchannel with hydrodynamic slip and thermal radiation. International Journal of Heat and Mass Transfer. 2016 Sep 1;100:89-97.
	\bibitem{43}Sohail M, Shah Z, Tassaddiq A, Kumam P, Roy P. Entropy generation in MHD Casson fluid flow with variable heat conductance and thermal conductivity over non-linear bi-directional stretching surface. Scientific Reports. 2020 Jul 27;10(1):12530.
	\bibitem{44}Estrada R, Ibáñez G, López A, Lastres O, Pantoja J, Reyes J. Analytical analysis of impacts of nanoparticle shapes and uncertainty in thermophysical properties on optimum operating conditions of MHD nanofluid flow in a microchannel filled with porous medium. Journal of Thermal Analysis and Calorimetry. 2024 Jan;149(1):265-98.
	\bibitem{45}Ghaderi E, Bijarchi M, Hannani SK, Nouri-Borujerdi A. Parametric and inverse analysis of flow inside an obstructed channel under the influence of magnetic field using physics informed neural networks. arXiv preprint arXiv:2404.17261. 2024 Apr 26.
\end{thebibliography}
\end{document}